\documentclass[12pt]{iopart}
\usepackage{iopams}
\usepackage{graphicx}
\usepackage{hyperref}

\begin{document}
\title{Mass inflation inside black holes revisited}
\author{Vyacheslav I. Dokuchaev}
\address{Institute for Nuclear Research of the Russian Academy of
Sciences, \\ Prospekt 60-letiya Oktyabrya 7a, Moscow 117312, Russia}

\begin{abstract}
The mass inflation phenomenon implies that black hole interiors are unstable due to a
back-reaction divergence of the perturbed black hole mass function at the Cauchy horizon. Weak point in the standard mass inflation calculations is in a fallacious using of the global Cauchy horizon as a place for the maximal growth of the back-reaction perturbations instead of the local inner apparent horizon. It is derived the new spherically symmetric back-reaction solution for two counter-streaming light-like fluxes near the inner apparent horizon of the charged black hole by taking into account its separation from the Cauchy horizon. In this solution the back-reaction perturbations of the background metric are truly the largest at the inner apparent horizon, but, nevertheless, remain small. The back reaction, additionally, removes the infinite blue-shift singularity at the inner apparent horizon and at the Cauchy horizon.
\end{abstract}



\section{Introduction}
The {\sl mass inflation} phenomenon, resulting in exponential growth of the perturbed black
hole mass function at the Cauchy horizon, was considered as a fatal instability of the interior
Kerr--Newman black hole solution with respect to the small perturbations (see, e.\,g.,
\cite{PoisIs89,PoisIs89b,PoisIs90,HamAvel10,Dong11} and an example of the bounded mass
inflation \cite{Brady}). The specific instability of the Cauchy horizon with respect to the
kink-mode perturbation was considered in the case of the Reissner--Nordstr\"om--(anti) de
Sitter black hole \cite{Maeda}. The problem of the Cauchy horizon instability has been probed
also by different numerical calculations (see, e.\,g., \cite{HodPiran98,HodPiran98b,Hong} and
references therein). The mass inflation problem in the case of the rotating black holes was
elaborated in \cite{Ori92}.

The standard mass inflation calculations \cite{PoisIs89,PoisIs89b,PoisIs90} are based on the
using of the generalized Dray--t'\,Hooft--Redmount (DTR) relation \cite{redm,drayth} in the
linear approximation of the Einstein equations near the perturbed inner horizon. However, the using of linear approximation to the DTR relation near horizons is a quite improper in view of the nonlinearity of the Einstein equations. This nonlinearity is especially crucial in the vicinity of the black hole horizons. 

An additional weak point in the standard mass inflation calculations is in a fallacious using
of the {\sl global} null Cauchy horizon as a place for the maximal growth of the back-reaction
perturbations instead of the {\sl local} inner apparent horizon, which is separated from the
Cauchy horizon. The maximal back-reaction perturbations inside the charged black hole take
place (besides the central singularity) at the local inner apparent horizon and not at the
separated global Cauchy horizon. This qualitative feature was missed in the previous mass
inflation calculations.

It is shown below that a back-reaction of two counter-streaming light-like fluxes result in
only the small corrections to the background metric near the local inner apparent horizon of
the charged black hole by taking into account its separation from the Cauchy horizon. This
implies the absence of the mass inflation and stability of the charged black hole interiors.

We use a (slow) quasi-stationary approximation, when both the inflowing and outflowing radial
energy fluxes of light-like particles are small, $\dot m_{\rm in}\ll1$ and $\dot m_{\rm
out}\ll1$. Respectively, the rate of black hole mass growth is also small (for more details
see, e.\,g., \cite{de2011,bde2012}). In finding the back-reaction in this approximation, we use
the Reissner--Nordstr\"om black hole solution as a background metric and retain in equations
only those perturbation terms, which are not higher, than the linear ones with respect to $\dot
m_{\rm in}$ and $\dot m_{\rm out}$. In other words, we calculate the back-reaction in the
linear perturbation approximation with respect to the small dimensionless energy flux
parameters, $\dot m_{\rm in}$ and $\dot m_{\rm out}$.

\section{Absence of mass inflation}

\subsection{Back-reaction metric in the $(v,r)$-frame}
\label{sectionvr}

A general space-time metric in the spherically symmetric case can be written in the form
\cite{Bondi64,LL}:
\begin{equation}
    \label{EF}
    ds^2 = e^{\lambda(v,r)}dv\left[e^{\nu(v,r)+\lambda(v,r)}dv - 2dr\right] -r^2d\Omega^2,
\end{equation}
where $d\Omega^2=d\theta^2+\sin^2\theta d\phi^2$ is a 2-sphere metric and $\lambda(v,r)$ and
$\nu(v,r)$ are two arbitrary functions of two coordinates $v$ and $r$. This is a coordinate
frame of the {\it Eddington--Finkelstein} type, related, in particular, with the ingoing null
geodesics $v=const$. In analogy with the Schwarzschild, Reissner--Nordstr\"om and charged
Vaidya metric, we use the following form for the metric function $\nu(v,r)$:
\begin{equation}
 \label{f}
 e^{\nu(v,r)} \equiv f\equiv1-\frac{2m(v,r)}{r}+\frac{e^2}{r^2},
\end{equation}
where $m(v,r)$ is a mass function and $e$ is an electric charge. In the special case of
$\lambda(v,r)=0$, this metric is the charged Vaidya solution. Respectively,  at
$\lambda(v,r)=0$ and $m(v,r)=const$, this metric is the Reissner--Nordstr\"om solution.

In the back-reaction calculations, we follow very closely to \cite{PoisIs90}, by using both the
ingoing and outgoing Vaidya metrics as perturbations for the background Reissner--Nordstr\"om
black hole metric. For the ingoing Vaidya metric in the $(v,r)$-frame we choose the mass
function $m(v,r)$ in the specific form:
\begin{equation}
m(v,r)=m_{\rm in}(v)=\left\{\begin{array}{ll}
\!m_0\left[1-\beta_0(v_0/v)^{p-1}\right] & \mbox{ at } v\geq v_0, \\
\!m_0(1-\beta_0) & \mbox{ at } v< v_0,
\end{array}
\right.
   \label{massin}
\end{equation}
with constants $v_0$, $m_0$, $\beta_0\ll1$ and $p\geq12$. Respectively, for the corresponding
outgoing Vaidya solution in the $(u,r)$-frame we choose quite a similar expression for the mass
function m(u,r):
\begin{equation}
m(u,r)=m_{\rm out}(u)=\left\{\begin{array}{ll}
\!m_0\left[1-\beta_1(u_0/u)^{q-1}\right] & \mbox{ at } u\geq u_0, \\
\!m_0(1-\beta_1) & \mbox{ at } u< u_0,
\end{array}
\right.
   \label{massout}
\end{equation}
with the additional constants $u_0$, $\beta_1\ll1$ and $q\geq12$.

\begin{figure}
\begin{center}
 \includegraphics[width=0.8\textwidth]{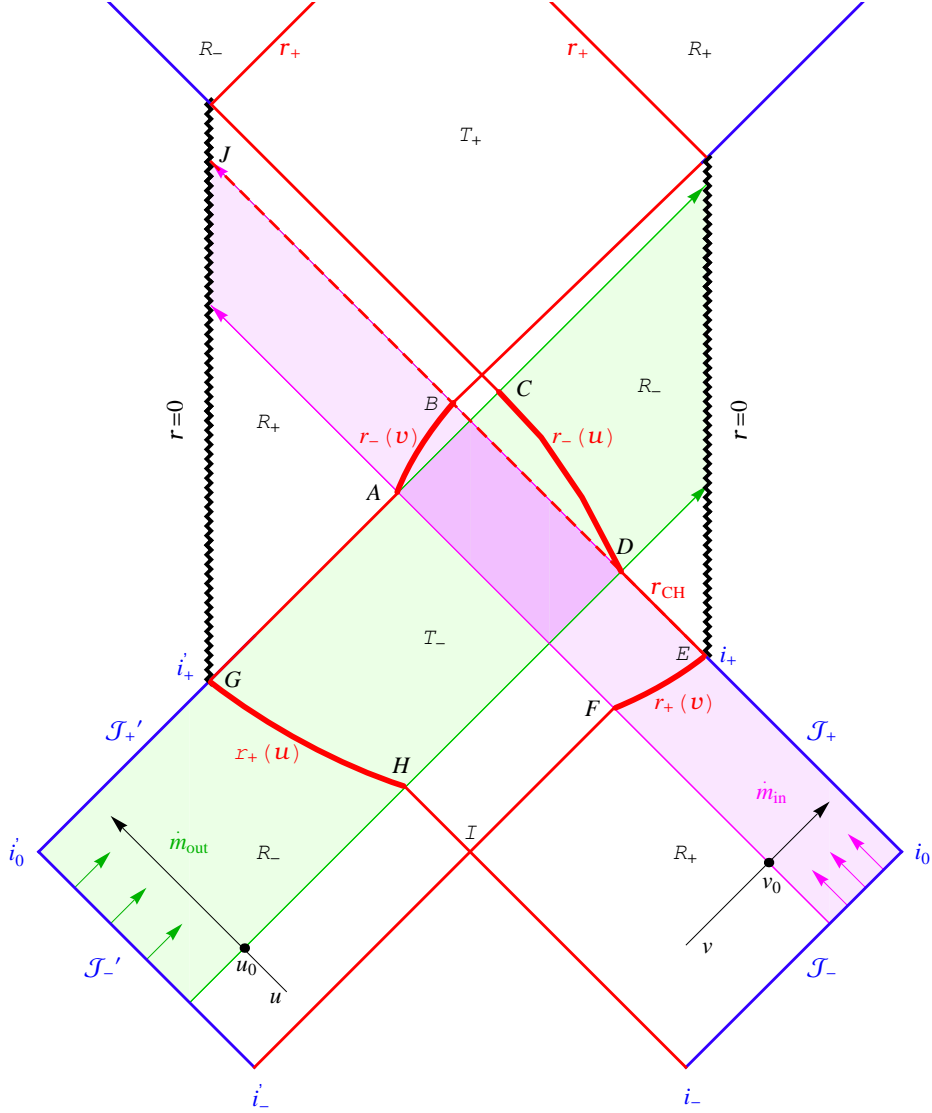}
 \end{center}
\caption{The Carter--Penrose diagram for the Reissner-Nordsr\"om black hole, perturbed by the
small inflowing, $\dot m_{\rm in}\ll1$, and outflowing, $\dot m_{\rm out}\ll1$, radial fluxes
of light-like particles, related with the corresponding ingoing and outgoing Vaidya solutions
outside the black hole. Inside the black hole, i.\,e., at $r<r_+(v)$ and $r<r_+(u)$, there is a
double filled overlapping region with two coexisting and oppositely directed fluxes. The global
Cauchy horizon $r_{\rm CH}$, which is defined by the null line $v-\infty$, is shown in part by
the null line $ED$ and further by the null dashed line $DBJ$. The perturbed metric deviates
from the linear sum of ingoing and outgoing Vaidya metrics (due to the nonlinearity of the
Einstein equations) in the space-time regions, corresponding to the all future-directed
light-cones in this overlapping region. The maximal perturbation of the black hole metric,
which is the most interesting for the discussed mass inflation problem. The maximal
blue-shift of the inward radiation, viewed by the free-moving observer, will take place just at
the part of the local inner apparent horizon $r_-(u)$, shown by the thick time-like curve $CD$
and not at the part of the global Cauchy horizon $r_{\rm CH}$, shown by the dashed null line
$BD$.}
 \label{CPdiagram}
\end{figure}

The global geometry of the Reissner--Nordstr\"om black hole, perturbed by both the ingoing and
outgoing radial null fluxes, is shown in Fig.~\ref{CPdiagram} by means of the Carter--Penrose
diagram. The metric functions $\lambda=0$ and $\nu=0$ at both the inner and outer apparent
horizons, $r=r_{\pm}(v)$ and $r=r_{\pm}(u)$, of the perturbed black hole. These local apparent
horizons are the boundaries between the $R$- and $T$-regions with the different metric
signatures, $(+,-,-,-)$ and $(-,+,-,-)$, respectively. The variable parts of these apparent
horizons are shown by the thick curves $AB$, $FE$, $GH$ and $CD$ in Fig.~\ref{CPdiagram}.

There is only one flux, respectively, inflowing $\dot m_{\rm in}$ or outflowing $\dot m_{\rm
out}$ in the filled but non-overlapping regions of the Carter--Penrose diagram in
Fig.~\ref{CPdiagram} for the perturbed metric. Solutions of the Einstein equations in these
non-overlapping regions are the corresponding ingoing and outgoing charged Vaidya metrics.
Meanwhile, both the inward and outward fluxes coexist in the overlapping region. It must be
stressed, that the overlapping region exists only inside the black hole, at $r<r_+(v)$ and
$r<r_+(u)$. In the overlapping region a corresponding solution of the Einstein equations
deviates from the Vaidya metric. In other words, the sum of two counter-streaming Vaidya
solutions is not the valid solution of the Einstein equations due to their nonlinearity.
Evidently, in the regions without any fluxes (see the non filled regions in
Fig.~\ref{CPdiagram} the standard Reissner--Nordstr\"om metric is realized, though with the
different values of the black hole mass before and after the passage of inflowing or outflowing
fluxes. The metric functions $\lambda=0$ and $\nu=0$ at both the inner and outer apparent
horizons, $r=r_{\pm}(v)$ and $r=r_{\pm}(u)$, of the perturbed black hole. These local apparent
horizons are the boundaries between the $R$- and $T$-regions with the different metric
signatures, $(+,-,-,-)$ and $(-,+,-,-)$, respectively.

A principal point is that the global Cauchy horizon $r=r_{\rm CH}$ is separated from the local
inner apparent horizon $r=r_-(v)$ in the case of the non-stationary (perturbed) metric. This
separation is clearly viewed in Fig.~\ref{CPdiagram}, where the global Cauchy horizon $r_{\rm
CH}$ is shown in part by the null line $ED$ and further by the dashed null line $DBJ$, while
the corresponding part of the inner apparent horizon $r=r_-(u)$ is shown by the time-like curve
$CD$. The maximal perturbation of the black hole metric, which is the most interesting for the
discussed mass inflation problem, and the maximal blue-shift of inward radiation, viewed by the
free-moving observer, will take place just at the $r_-(u)$ (the time-like curve $CD$) and not
at the $r_{\rm CH}$ (the null line $BD$).

The singular behavior of metric functions $\lambda$ and $\nu$ at the horizons is especially
crucial when the double-null (u,v)-frame is used near horizons. In particular, it was
demonstrated the ill-posedness of a double null ``free'' evolution scheme in numerical
calculations for the Einstein equations, when constraints are imposed only at the boundaries,
and all fields are propagated by means of the evolution equations \cite{GundlachPullin}. This
ill-posedness of the ``free'' evolution numerical scheme results in the artificial
exponentially growing mode.

To resolve a back-reaction problem for the perturbed spherically symmetric black hole, the two
sought functions $\nu$ and $\lambda$ in the general metric (\ref{EF}) must be calculated by
using the Einstein equations with the appropriately chosen perturbations. We follow below
closely to E. Poisson and W. Israel \cite{PoisIs90} by writing the Einstein equations in the
form
\begin{equation}
 G_{\alpha\beta}=8\pi(E_{\alpha\beta}+T_{\alpha\beta}),
  \label{Einstein}
\end{equation}
where the Maxwellian contribution to the energy-momentum tensor from the black hole electric
charge $e$ is
\begin{equation}
 E^\alpha_{\beta}=\frac{e^2}{8\pi r^4}\,{\rm diag}(1,1,-1,-1)
 \label{Maxwellian}
\end{equation}
and $T_{\alpha\beta}$ is a perturbation energy-momentum tensor.

The nonzero components of the Einstein tensor $G_{\alpha\beta}$ in the Eddington--Finkelstein
frame (\ref{EF}) are
\begin{eqnarray}
 \label{EinsteinG}
  G_0^{0} &=& -e^\nu\left(\frac{1}{r^2}+\frac{\nu'}{r}\right)
 +\frac{1}{r^2}, \label{EG00} \\
  G_0^{1} &=& \frac{e^\nu}{r}\,\dot\nu, \label{EG01} \\
  G_1^{0}  &=&  -2\frac{e^{-\lambda}}{r}\lambda',  \label{EG10} \\
  G_1^{1}  &=&   G_0^{0}-\frac{2\lambda'}{r}\,e^\nu, \label{EG11} \\
  G_2^{2} &=& G_3^{3}
 = -e^\nu\left(\lambda''+\frac{\nu''}{2}\right)-e^{-\lambda}\dot{\lambda}' \nonumber \\
 && -e^\nu\left(\lambda'^2 +\frac{\nu'^2}{2}+\frac{\lambda'+\nu'}{r}
  +\frac32\lambda'\nu'\right),    \label{EG22}
\end{eqnarray}
where dot \ $^.\!=\partial/\partial v$, and prime \ $'\!= \partial/\partial r$. The
cor\-res\-ponding Einstein equations are related by the Bianchi identity. Therefore, one of
them is not independent, e.\,g., equation, related with the component $G_2^{2}$ from
(\ref{EG22}). As a background metric we consider the Reissner--Nordsr\"om black hole metric,
which is an exact electro-vacuum solution of the Einstein equations (\ref{Einstein}) with
$T_{\alpha\beta}=0$. We calculate in the following the back-reaction on the background
Reissner--Nordsr\"om black hole metric of both the inflowing and outflowing radial fluxes of
light-like particles, described by the perturbation energy-momentum tensor
\begin{equation}
 \label{tensorinout}
 T_{\alpha\beta}=\rho_{\rm in} l_\alpha l_\beta+\rho_{\rm out} n_\alpha n_\beta
\end{equation}
with, respectively, the energy influx $\dot m_{\rm in}=4\pi r^2\rho_{\rm in}$ and outflux $\dot
m_{\rm out}=4\pi r^2\rho_{\rm out}$ and the radial null vectors
\begin{eqnarray}
 \label{l}
 l^\mu &=&  (l^v,l^r,l^\theta,l^\phi) = (0,-1,0,0), \\
 n^\mu &=&  (n^v,n^r,n^\theta,n^\phi)= (1,\frac{1}{2}fe^{\lambda},0,0),
 \label{n}
\end{eqnarray}
where $l^2=n^2=0$. Both the outer horizon $r_+$ and the inner horizon $r_-$ in the used
quasi-stationary approximation (with a linear accuracy in $\dot m_{\rm in}\ll1$ and $\dot
m_{\rm out}\ll1$) are solutions of the equation $f=0$ or, formally,  $r_\pm=
m(v,r)\pm\sqrt{m(v,r)^2-e^2}$ at $e^2\leq m(v,r)^2$. Note that these horizons are,
respectively,  the {\sl inner and outer apparent horizons} for a non-stationary metric
\cite{BoothMartin}.

The perturbation energy-momentum tensor components in the $(v,r)$-frame (\ref{EF}) are
\begin{eqnarray}
 \label{T00}
 T^0_0 &=& \frac{1}{2}fe^{2\lambda}\rho_{\rm out}=-T^1_1 , \\
 \label{T01}
 T^1_0 &=& e^{\lambda}(-\rho_{\rm in}+ \frac{1}{4}f^2e^{2\lambda}\rho_{\rm out}), \\
 \label{T10}
 T^0_1 &=& -e^{\lambda}\rho_{\rm out}, \\
 \label{T22}
 T^2_2 &=& T^3_3=0.
\end{eqnarray}
We calculate in the following the back-reaction solution of the Einstein equations near
horizons, at $|r-r_{\pm}|/r_{\pm}\ll1$. In the quasi-stationary approximation the both fluxes
are nearly constant in the vicinity of horizons, $\dot m_{\rm in}=const\ll1$ and $\dot m_{\rm
out}=const\ll1$, respectively. From equation (\ref{EG10}) we get near horizons
\begin{equation}
  e^{\lambda}\approx\left\{
  \begin{array}{ll}
  1 & \mbox{~in regions without ouward flux}, \\
 \left(r/r_\pm\right)^{\dot m_{\rm out}} & \mbox{~in regions with outward flux}.
  \end{array} \right.
  \label{lambda4}
\end{equation}
Respectively, from equations (\ref{EG00}) and (\ref{EG11}) we  get
\begin{eqnarray}
 \label{mr}
 m' &=& 4\pi r^2 T^0_0 = \frac{1}{2}\dot m_{\rm out}fe^{2\lambda}, \\
 \dot m &=& -4\pi r^2 T^1_0
 = e^{\lambda}(\dot m_{\rm in}- \frac{1}{4}f^2e^{2\lambda}\dot m_{\rm out}).
 \label{mV}
 \end{eqnarray}
Near horizons, where $f\ll1$ and $\lambda\ll1$, we have
\begin{equation}
 \label{mrV}
m'\approx \frac{1}{2}\,\dot m_{\rm out} f, \quad \dot m \approx\dot m_{\rm in}.
 \end{equation}
The mass function $m(v,r)$ strongly depends on the coordinate $r$ but weakly on the coordinate
$v$ near $r=r_\pm$ in the used quasi-stationary approximation. For this reason, it is credible
to adopt a factorization for the mass function near horizons (see also \cite{de2011}):
$m(v,r)=m(v)\mu(x)$ with the dimensionless coordinate $x=r/m_0$ and with the ``mass''
$m(v)=m_{\rm in}(v)$, where $m_{\rm in}(v)$ is from equation (\ref{massin}). The function
$m(v)$ is weakly growing with $v$, i.\,e., $dm(v)/dv=\dot m_{\rm in}\ll1$ and, therefore,
$m(v)\approx m_0=const$.

To solve the first nonlinear equation in (\ref{mrV}) near the outer and inner apparent horizons
$x=x_\pm$, we define the black hole extreme parameter $\epsilon=\sqrt{1-e^2/m_0^2}$ and the new
variable
\begin{equation}
\label{deltapm}
 \delta_\pm(x)\equiv x-x_\pm =x-[\mu(x)\pm \sqrt{\mu(x)^2-(1-\epsilon^2)}],
\end{equation}
The horizons $x=x_{\pm}$ are solutions of equation $\delta_{\pm}(x)=0$. Near horizons, at
$|\delta_{\pm}|\ll1$, we have
\begin{equation}
 \label{fhorpmVr}
 f \approx
 \pm2\frac{\sqrt{\mu(x)^2-(1-\epsilon^2)}}{x_\pm^2}\delta_\pm
 \approx \pm\frac{2\epsilon}{(1\pm\epsilon)^2}\delta_\pm.
 \end{equation}
Now the left equation in (\ref{mrV}) is written as
\begin{equation}
 \label{mrVhor}
\frac{d\mu}{dx}\approx \pm\frac{\epsilon\dot m_{\rm out}}{(1\pm\epsilon)^2}(x-x_\pm).
 \end{equation}
Solution of this equation in the non-overlapping region, where there is only influx, is
$\mu(x)=1$. Evidently, this solution coincides with the corresponding ingoing Vaidya metric in the non-overlapping region in the vicinity of the null line $HD$ in Fig.~\ref{CPdiagram}.

Meantime, solution of the same equation (\ref{mrVhor}) in the vicinity of the inner apparent
horizon, at $|x-x_-|\ll1$, in the overlapping region, where there are both the inward and
outward fluxes, is
\begin{equation}
 \mu(x)\approx 1-\frac{\epsilon\dot m_{\rm
out}}{(1-\epsilon)^2}\frac{(x-x_-)^2}{2}.
  \label{muxRN}
\end{equation}
The resulting metric deviates from the Vaidya solution in the overlapping region near the inner
apparent horizon $x=x_-$ (in the vicinity of the curve $HD$ in Fig.~\ref{CPdiagram}). The
integration constant in (\ref{muxRN}) is $1$ due to continuity relation, i.\,e., this solution
must coincide with the Vaidya solution at the inner horizon $x=x_-$ on the border between the
overlapping and non-overlapping regions (at the point $D$ in Fig.~\ref{CPdiagram}).

At the same time, the solution of equation (\ref{mrVhor}) near both the inner and outer
apparent horizons, at $x=x_\pm$, in the non-overlapping regions, where there is only outward
flux $\dot m_{\rm out}$ is the outgoing Vaidya metric:
\begin{equation}
 \mu(x)\approx 1\pm\frac{\epsilon\dot m_{\rm
out}}{(1\pm\epsilon)^2}\frac{(x-x_\pm)^2}{2}.
  \label{muxRNvaidya}
\end{equation}
This outgoing Vaidya solution, however, is presented here in the ingoing (v,r)-frame, which is
non-optimal in the presence of outward flux $\dot m_{\rm out}$. In the outgoing (u,r)-frame the
mass function $m(u,r)$ would have the vice-versa behavior.

In general, there are two variable branches of the inner apparent horizon, $r=r_-(v)$ and
$r=r_-(u)$, shown by the thick curves $AB$ and $CD$ in Fig.~\ref{CPdiagram}. The back-reaction
mass function $m(v,r)\approx m_0\mu(x)$ from (\ref{muxRN}) is finite at the both inner apparent
horizons $AB$ and $CD$ without any indication of the mass inflation.

\subsection{Back-reaction metric in the $(t,r)$-frame}

An alternative approach is in using of the general spherically-symmetric metric in the
Schwarzschild-like $(t,r)$-frame \cite{LL}:
\begin{equation}
 \label{spherical}
 ds^2=e^{\eta(t,r)} dt^2-e^{\sigma(t,r)}dr^2-r^2d\Omega^2
 \end{equation}
with two arbitrary functions, $\eta(t,r)$ and $\sigma(t,r)$. For application to the
back-reaction problem of the accreting matter on the charged black hole, we define,
additionally, two metric functions, $f_0(t,r)$ and $f_1(t,r)$ or, equivalently, two {\sl mass
functions}, $m_0(t,r)$ and $m_1(t,r)$:
\begin{eqnarray}
 e^{\eta(t,r)} &\equiv& f_0=1-\frac{2m_0(t,r)}{r}+\frac{e^2}{r^2},  \label{m0}
 \\
 e^{-\sigma(t,r)} &\equiv& f_1=1-\frac{2m_1(t,r)}{r}+\frac{e^2}{r^2}. \label{m1}
\end{eqnarray}
The nonzero components of the Einstein tensor $G_{\alpha\beta}$ in the $(t,r)$-frame
(\ref{spherical}) have the following form \cite{LL}:
\begin{eqnarray}
 \label{G01}
  G^1_0 &=& -e^{-\sigma}\frac{\dot\sigma}{r}, \\
 \label{G00}
  G^0_0 &=& -e^{-\sigma}\left(\frac{1}{r^2}-
 \frac{\sigma'}{r}\right)+\frac{1}{r^2}, \\
 \label{G11}
  G^1_1 &=& -e^{-\sigma}\left(\frac{1}{r^2}+
 \frac{\eta'}{r}\right)+\frac{1}{r^2}, \\
 \label{G22}
  G^2_2=G^3_3 &=& \frac{e^{-\eta}}{2}
\left[\ddot\sigma+\frac{\dot\sigma}{2}\left(\dot\sigma-\dot\eta\right)\right]
 -\frac{e^{-\sigma}}{2}
 \left[\eta''+(\eta'-\sigma')\left(\frac{\eta'}{2}+\frac{1}{x}\right)\right]\!.
\end{eqnarray}
For the inward and outward null vectors in the energy-momentum tensor (\ref{tensorinout}) we
choose now
\begin{equation}
 \label{ltr}
 l_a = (-1,-\frac{1}{\sqrt{f_0f_1}},0,0), \quad
 n_a = (1,-\frac{1}{\sqrt{f_0f_1}},0,0).
\end{equation}
The corresponding perturbation energy-momentum tensor (\ref{tensorinout}) is now
\begin{eqnarray}
 \label{T00tr}
 T^0_0 &=& \frac{1}{f_0}(\rho_{\rm in}+\rho_{\rm out})=-T^1_1 , \\
 \label{T01t}
 T^1_0 &=& \sqrt{\frac{f_1}{f_0}}(\rho_{\rm out}-\rho_{\rm in}), \\
 \label{T10tr}
 T^2_2 &=& T^3_3=0.
\end{eqnarray}
From equations (\ref{Einstein}), (\ref{G01}), (\ref{G00}) and (\ref{G00}) with perturbation
energy-momentum tensor from  (\ref{T00tr}) and (\ref{T01t}) we get the requested back reaction
equations for the mass  functions $m_0(t,r)$ and $m_1(t,r)$ in the vicinity of the apparent
horizons at $|r-r_\pm|/r_\pm\ll1$, where $f_0\ll1$ and $f_1\ll1$:
\begin{eqnarray}
 \label{dotm1}
 \dot m_1 &=& \frac{\partial m_1}{\partial  t}=\sqrt{\frac{f_1}{f_0}}
 (\dot m_{\rm in}-\dot m_{\rm out})\ll1, \\
 m'_1 &=& \frac{\partial m_1}{\partial  x}
 \approx\frac{\dot m_{\rm in}+\dot m_{\rm out}}{f_0}\approx-m'_0.
 \label{primem}
\end{eqnarray}
In the quasi-stationary approximation the both mass functions, $m_0(t,r)$ and $m_1(t,r)$,
strongly depend on the radial coordinate $r$, but very weakly on the time coordinate $t$. For this
reason, as in the previous Section~\ref{sectionvr}, we adopt near horizons a factorization for
the mass functions: $m_0(t,r)=m(t)\mu_0(x)$ and $m_1(t,r)=m(t)\mu_1(x)$ with the dimensionless
coordinate $x=r/m_0$ and with a weakly growing ``mass'' $m(t)$. We use also the variable
$\delta_\pm$ from equation (\ref{deltapm}) to solve the nonlinear equation (\ref{primem}) in
the vicinity of the inner and outer apparent horizons. Again, as in the
Section~\ref{sectionvr}, the outer and inner apparent horizons at $x=x_{\pm}$ are solutions of
equation $\delta_{\pm}(x)=0$, where $\delta_{\pm}(x)$ is defined by equation (\ref{deltapm}). Now, near the apparent horizons, at $|x-x_\pm|\ll1$, where
$|\delta_{\pm}|\ll1$, we have
\begin{equation}
 \label{f1horpm}
 f_0 \approx
 \pm2\frac{\sqrt{\mu_0(x)^2-(1-\epsilon^2)}}{x_\pm^2}\delta_\pm
 \approx \pm\frac{2\epsilon}{(1\pm\epsilon)^2}\delta_\pm.
 \end{equation}
Respectively, equation (\ref{primem}) is written near the apparent horizons as
\begin{equation}
 \mu'_0 =\frac{d\mu_0}{dx}
 \approx \mp\frac{(1\pm\epsilon)^2}{2\epsilon}
 \frac{\dot m_{\rm in}+\dot m_{\rm out}}{\delta_\pm(x)}\approx-\mu'_1. \label{primem0}
\end{equation}
The resulting solutions for mass functions near the apparent horizons are
\begin{eqnarray}
 \mu_0(\delta_\pm) &\approx& \mu_\pm\mp\frac{(1\pm\epsilon)^2}{2\epsilon}\,\dot m
 \log{\left|1+\frac{2\epsilon^2}{(1\pm\epsilon)^3\dot m}\,\delta_{\pm}\right|},  \label{muRN0}
 \\
 \mu_1(\delta_\pm)&\approx&2\mu_\pm-\mu_0(\delta_\pm).
  \label{muRN1}
\end{eqnarray}
Here
\begin{equation}
 \dot m=\left\{
  \begin{array}{ll}
  \!0 & \mbox{~in regions without fluxes}, \\
 \!\dot m_{\rm in} & \mbox{~in non-overlapping regions with influx}, \\
 \!\dot m_{\rm in}+\dot m_{\rm out}& \mbox{~in overlapping region}, \\
  \!\dot m_{\rm out} & \mbox{~in non-overlapping regions with outflux}, \\
  \end{array} \right.
  \label{fluxinout}
\end{equation}
and the values of mass functions at the horizons, respectively, $\mu_+\equiv
\mu_0(x_+)=\mu_1(x_+)$ and $\mu_-\equiv\mu_0(x_-)=\mu_1(x_-)$ are
\begin{equation}
 \mu_{\pm}\approx 1\mp\frac{(1\pm\epsilon)^2}{2\epsilon}\,\dot m
 \log{\left|\dot m\right|}.
 \label{mu+2}
\end{equation}
It can be seen in Fig.~\ref{CPdiagram}, that all the variable parts of the horizons, i.\,e.,
the apparent horizons $r_\pm(u)$ and $r_\pm(v)$, are placed in the non-overlapping regions of
the Carter--Penrose diagram, where there is only one flux, inward or outward (curves $AB$,
$CD$, $EF$ and $GH$). At the same time, the constant parts of the inner and outer horizons are
placed either at the borders of the non-overlapping regions with the zero fluxes (null lines
$AG$ and $DE$), corresponding to the static black hole with a mass $m_0(1-\epsilon)$, or in the
regions without any flux (null lines $FI$ and $HI$), corresponding to the static black holes
with masses $m_0(1-\epsilon)(1-\beta_0)$ and $m_0(1-\epsilon)(1-\beta_1)$. Note, that the null
line $DE$, which is a constant part of the inner apparent horizon $r_-(u)$ is identical to the
part of Cauchy horizon $r_{\rm CH}$) of the global metric. The most interesting for us is the
time-like curve $CD$ of the inner apparent horizon $r_-(u)$ of the perturbed black hole. The
mass functions $\mu_0$ and $\mu_1$ are equal at $r=r_-(u)$ with a corresponding value
\begin{equation}
 \mu_-\approx 1+\frac{(1-\epsilon)^2}{2\epsilon}\,\dot m_{\rm out}
 \log{\left|\dot m_{\rm out}\right|}<1.
 \label{mu-}
\end{equation}
according to equation (\ref{mu+2}).

\begin{figure}
\begin{center}
  \includegraphics[width=0.8\textwidth]{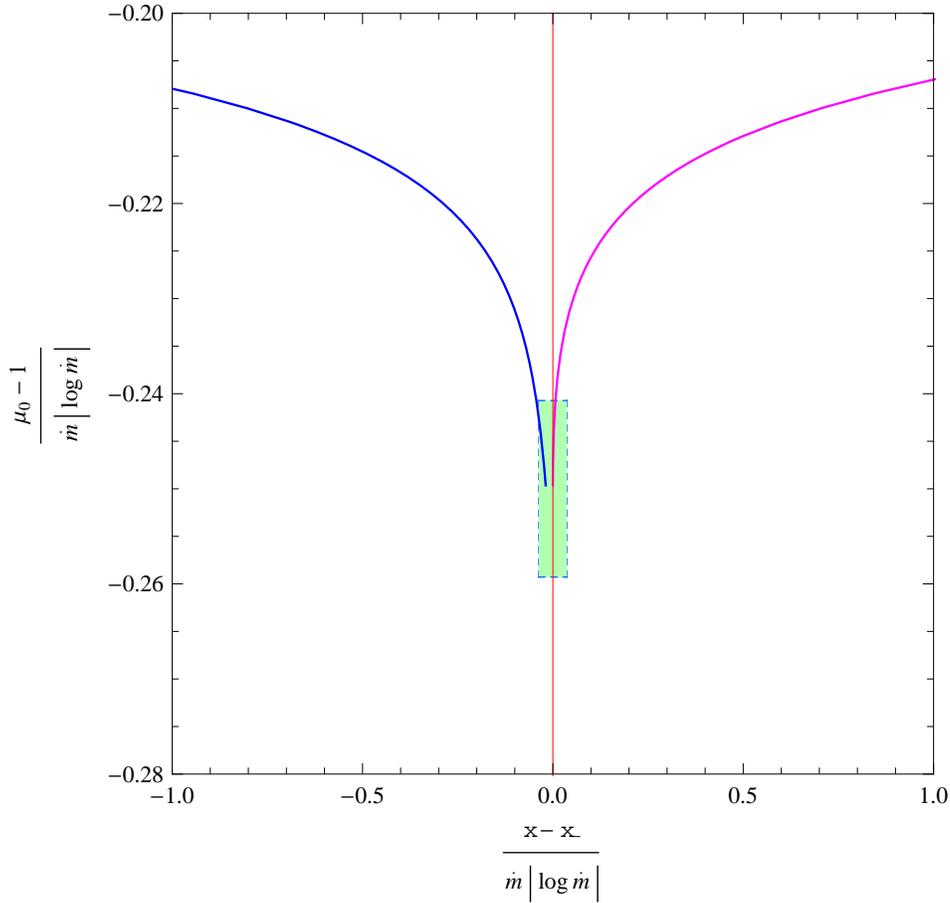}
 \end{center}
\caption{The mass function $\mu_0(x)$ from equation (\ref{muRN0}) for the black hole with the
charge $e=\sqrt{3}/2$ and flux $\dot m=\dot m_{\rm out}=10^{-12}$ near the inner apparent
horizon $x=x_-$, modified by the back-reaction. Inside the filled box the used linear
perturbation approximation in $\dot m$ is insufficient for the exact determination of the mass
function $\mu_0(x)$. In the used approximation, the value of the $\mu_0(x)$ at the $x=x_-$ is
finite and defined by equation (\ref{mu-}). Perturbations of the mass functions $\mu_0(x)$ and
$\mu_1(x)$ remain small at the inner apparent horizon,
$|\mu_0(x_-)-1|=|\mu_1(x_-)-1|\propto\dot m|\log{\dot m}|\ll1$.}
 \label{mu0minus}
\end{figure}

Solution (\ref{muRN0}) for the mass function $\mu_0(x)$ near horizons may be written in the
form of the inverse function $x=x(\mu_0)$, where
\begin{equation}
 \label{inverse}
 x(\mu_0)\approx \mu_0\pm\!\sqrt{\mu_0^2\!-\!(1\!-\!\epsilon^2)}\!
 +\!\frac{(1\pm\epsilon)^3\dot m}{2\epsilon^2}
 \left\{\!\pm  \exp\left[\mp\frac{2\epsilon(\mu_0\!
 -\!\mu_\pm)}{(1\pm\epsilon)^2\dot m}\right]\!-\!1\!\right\}\!.
  \end{equation}
Here, the signs $\pm$ and $\mp$ are related to the different brunches of the function
$x(\mu_0)$: the first branch for $x\leq x_-$ and the second one for $x\geq x_-$. These branches
are shown in Fig.~\ref{mu0minus} for the case of $\mu_0(x)$ near the inner apparent horizon,
corresponding to the curve $CD$ in Fig.~\ref{CPdiagram}, at $|x-x_-|\ll1$. The first branch is
at $x\leq x_-$ and the second one is at $x\geq x_-$.

With the chosen approximations, solutions (\ref{muRN0}), (\ref{muRN1}) and (\ref{inverse}) are
valid only in the narrow region $\delta_{\rm min} \equiv [(1\pm\epsilon)^3/\epsilon^2]\dot m
<|\delta_\pm|\ll1$ near the apparent horizons. In these solutions we retain only the leading
perturbation terms $\sim\dot m|\log\dot m|\ll1$ and neglect the much more smaller
contributions, $\sim\dot m\ll1$. The used linear perturbation approximation with respect to the
small parameter $\dot m\ll1$ would be insufficient for calculations of the mass functions at
$|\delta_\pm|<|\delta_{\rm min}|$, where the quadratic perturbation terms $\propto \dot m^2$
and the higher orders ones must be taking into account. See in Fig.~\ref{mu0minus} the mass
function $\mu_0(x)$ from equation (\ref{muRN0}) near the inner apparent horizon $x_-$. Note
also, that divergence in (\ref{mu+2}) in the extreme limit at $\epsilon\to0$ is related with a
general instability of the perturbed extreme black hole (for some details see, e.\,g.,
\cite{de2011}).

The derived solutions (\ref{muRN0}) and (\ref{muRN1}) demonstrate that the back-reaction
corrections to the mass functions $m_0(t,r)$ and $m_1(t,r)$ are small near and at the inner
apparent horizon of the non extreme black hole. Namely, the relative disturbance of the inner
apparent horizon is small, of the order of $\dot m|\log{\dot m}|\ll1$. Therefore, the mass
inflation phenomenon is absent at least in the used quasi-stationary approximation.

\section{Absence of the infinite blue-shift singularity}

We calculate the blue-shift of the inward radiation viewed by a free-moving observer,
traversing the non-stationary local apparent horizon $r_-$ or the global Cauchy horizon $r_{\rm
CH}$ of the perturbed charged black hole. The new feature in this calculation, with respect to
a previous similar blue-shift calculation in \cite{PoisIs90}, is in taking into account, at
fist, the non-stationarity of the black hole due to the presence of the influx of light-like
particles and, at second, the additional perturbation, produced by the observer himself.

The appropriate place for a possible large blue-shift is a border between the $T_-$-region and
$R_-$-region inside the charged black hole. In the discussed model, this border composed of the
part of the inner apparent horizon $r_-(u)$ and the part of Cauchy horizon $r_{\rm CH}$, shown,
respectively, by the time-like curve $CD$ and the null line $DE$ in Fig.~\ref{CPdiagram}.

Following closely to \cite{PoisIs90}, we use, at first, the ingoing Vaidya solution as a
background metric with a mass function $m(v)$ from equation (\ref{massin}). The influx of
light-like particles in the ingoing Vaidya metric is described by the energy-momentum tensor
$T^{\rm in}_{\alpha\beta}=\rho_{\rm in} l_\alpha l_\beta$, where $l_\alpha=-\partial_\alpha v$
and $4\pi r^2\rho_{\rm in}=dm_{\rm in}/dv$. The requested energy density, measured by a
free-moving observer with four-velocity $u^\alpha$, is
\begin{equation}
 \label{rhoobs}
 \rho_{\rm obs}=T_{\alpha\beta}u^\alpha u^\beta=
 \frac{(l_\alpha u^\alpha)^2}{4\pi r^2}\frac{dm_{\rm in}}{dv},
\end{equation}
where $dm_{\rm in}/dv$ may be calculated, e.\,g., from (\ref{massin}). By using the
four-velocity normalization $u^\alpha u_\alpha=1$, we get the trajectory equations for the
free-moving observers of two dissimilar kinds (starting for simplicity somewhere inside the $T_-$-region in Fig.~\ref{CPdiagram}):
\begin{equation}
 \label{v}
 \dot v_{1,2}=f^{-1}[\dot r\pm[\dot r^2+f)^{1/2}],
\end{equation}
where overdot denotes differentiation with respect to the observer's proper time. Observer of the first-kind (we call him the ``red-shift observer'') is traversing the border between the $T_-$-region and $R_+$-region, while the second-kind observer (we call him the ``blue-shift observer'') is traversing the border between the $T_-$-region and $R_-$-region (see
Fig.~\ref{CPdiagram}).

For the red-shift observer in the $T_-$-region and near the apparent horizon $r_-$, where $\dot r<0$, $0<-f\ll1$ or $(r-r_-)/r_-\ll1$, we would have from (\ref{v}) the trajectory equation
\begin{equation}
 \label{v1}
 \dot v_1=f^{-1}[\dot r+[\dot r^2+f)^{1/2}]\approx-\frac{2}{\dot r}.
\end{equation}
It is evidently seen in Fig.~\ref{CPdiagram} that any red-shift observer traverses the
non-stationary part of the inner apparent horizon $r_-$, shown by the space-like curve $AB$,
at the finite values of $v$. Respectively, the value of $\dot v_1$ is finite at $r=r_-$
according to equation (\ref{v1}). In result, the red-shift observer will see the finite
gravitational red-shift of the influx $\dot m_{\rm in}$ at the inner apparent horizon
$r=r_-(v)$.

Meantime, for the blue-shift observer, in the same $T_-$-region and near the apparent or Cauchy
horizons, we would have from (\ref{v}) quite the different trajectory equation
\begin{equation}
 \label{v2}
 \dot v_2=f^{-1}[\dot r-[\dot r^2+f)^{1/2}]\approx\frac{2\dot r}{f}
\end{equation}
The blue-shift observer may traverse either the {\sl non-stationary} part of the inner apparent
horizon $r_-(u)$, shown by the time-like curve $CD$ in Fig.~\ref{CPdiagram}, or the stationary
part of the inner apparent horizon, which is the part of the global Cauchy horizon, $r_-=r_{\rm
CH}=m_0(1-\epsilon)=const$, shown by the null line $DE$ in Fig.~\ref{CPdiagram}.

The inner apparent horizon $r_-(u)$ is placed in the space-time beyond the null-line $v=\infty$
because $r_-(u)\leq r_{\rm CH}$. For this reason we need to use the $u,r$-frame for the
discussed ingoing Vaidya solution to solve the trajectory equation for the blue-shift observer,
traversing  the inner apparent horizon $r_-(u)$. To circumvent this difficulty we note, that
the blue-shift observer in the $(v,r)$-frame corresponds to the red-shift one in the $(u,r)$-frame. This means that a blue-shift observer will see the finite gravitational red-shift of the
outflux $\dot m_{\rm out}$, traversing the inner apparent horizon  $r=r_-(u)$, for the similar
reasons, as the discussed earlier, the red-shift observer, traversing the inner apparent
horizon $r=r_-(v)$.

To calculate the values of the possible blue-shift or red-shift, viewed by observers traversing
the inner apparent horizons $r_-$, it is needed to use the geodesic equations in the perturbed Vaidya metric, which is beyond the scope of this paper. For the sake of this paper it is enough to establish that these values are finite.

Quite the contrary, the blue-shift observer traverse the Cauchy horizon $r_{\rm CH}$ at the
infinite value of coordinate $v$. Therefore, $(l_\alpha u^\alpha)^2=\dot v^2\propto f^{-2}$, is
infinite at the Cauchy horizon for these observers. We integrate now equation (\ref{v}) near
the Cauchy horizon $r_{\rm CH}$ (i.\,e., in the vicinity of the null line $ED$ in
Fig.~\ref{CPdiagram}). Differentiation of the metric function $f$ from (\ref{f}) with respect
to variables $v$ and $r$ with a mass function $m(v)$ for the ingoing Vaidya metric yields the
trajectory equation for a free moving second kind observer in the $T_-$-region near the Cauchy
horizon:
\begin{equation}
   \label{dfdv}
 \frac{df}{dv}\approx -2\kappa_-\frac{dr}{dv} - \frac{2}{r_{\rm CH}}\frac{dm}{dv},
\end{equation}
where
\begin{equation}
   \label{kappa-}
 \kappa_-=\frac{\epsilon}{m_0(1-\epsilon)^2}
\end{equation}
is a surface gravity at the Cauchy horizon $r_{\rm CH}=m_0(1-\epsilon)$. Substituting $dv/dr$
from (\ref{v2}) into equation (\ref{dfdv}), we obtain
\begin{equation}
   \label{dfdv2}
 \frac{df}{dv} + \kappa_- f\approx- \frac{2}{m_0(1-\epsilon)}\frac{dm}{dv},
\end{equation}
It is crucial that a term with $dm/dv$ in (\ref{dfdv2}), related with the black hole
non-stationarity, was missed in previous calculations of the infinite blue-shift at the Cauchy
horizon.

Solution of equation (\ref{dfdv2}) with $m(v)=m_{\rm in}(v)$ from (\ref{massin}) is
\begin{equation}
   \label{ff}
f\approx -2(-1)^{p}(p-1)\frac{m_0\beta_0}{v_0}(k_-v_0)^{p-1}e^{-k_-v}\,\,\Gamma(1-p,-k_-v),
\end{equation}
where
\begin{equation}
   \label{Gamma}
\Gamma(a,x)=\int_x^\infty t^{a-1}e^{-t}dt
\end{equation}
is an incomplete Gamma-function. The asymptotic form of this solution is
\begin{eqnarray}
   \label{ff2}
 f\approx-2\beta_0\frac{p-1}{1-\epsilon}\left(\frac{v_0}{v}\right)^p\propto v^{-p}
 \quad \mbox{at } \quad v\gg v_0.
\end{eqnarray}
We neglect in this expression the terms of the order of $1/v^{p+1}$ and higher. In result,
$\rho_{\rm obs}\propto v^p$ at $v\gg v_0$, indicating the power-law blue-shift divergence at
the Cauchy horizon at $v\to\infty$. E. Poisson and W. Israel obtained a much more stronger
exponential divergence for $\rho_{\rm obs}$ (see, e.\,g., Eqs.~(B7) and (B8) in
\cite{PoisIs90}) by neglecting the black hole non-stationarity, i.\,e., by putting $dm_{\rm
in}/dv=0$ in equation (\ref{dfdv2}).

The derived blue-shift at the Cauchy horizon $r=r_{\rm CH}$ is infinite for quite a formal
reason: it was considered the static black hole metric with $r_{\rm CH}=const$ with the
ignorance of the inevitable perturbation of the black hole metric, produced by the moving
observer. This perturbation, even if extremely small, will separate the local inner apparent
horizon $r_-(u)$ from the global Cauchy horizon. Physically, this metric perturbation from the
free moving observer may be qualitatively modeled by the discussed outflux $\dot m_{\rm
out}\ll1$.  In result, the blue-shift observer will traverse the local inner apparent horizon
$r_-(u)$, viewing only the finite blue-shift of the influx. Note that the blue-shift observer will not see any considerable blue-shift of ingoing radiation $\dot m_{\rm in}$ at the dashed part of the Cauchy horizon $r=r_{\rm CH}$ (null line $BD$ in Fig.~\ref{CPdiagram}).

\section{Conclusion}

Solution of the perturbation back reaction problem for the non extreme Reissner-Nordstr\"om
black hole reveals no indication of the mass inflation by taking into account the separation of
the inner apparent horizon from the Cauchy horizon in the (slow) quasi-stationary accretion
approximation. This separation was missed in the previous calculations of the mass inflation
phenomenon.

The relative back-reaction corrections to the perturbed metric in the (v,r)-frame and
(t,r)-frame at the finite distance from both the inner and outer apparent horizons,
$|r-r_\pm|/r_\pm\sim1$, appear to be of the order of small accretion rate, $\dot m\ll1$, which
is a small dimensionless energy flux parameter. At the same tine, near and at the apparent
horizons, at $|r-r_\pm|/r_\pm\ll1$, the relative back-reaction corrections to the black hole
metric are the largest, but still remain the small, of the order of $\dot m\log{\dot m}\ll1$.
This means the absence of mass inflation phenomenon inside the charged black hole.
Additionally, it shown that a back reaction removes the infinite blue-shift singularity at the
inner apparent horizon.

There are a lot of limitations to the validity of the derived back-reaction corrections. The most vulnerable approximation is the slow accretion rate. It must be stressed also that semi-classical effects may influence the behavior of matter in the vicinity of the Cauchy horizon. Note also that the left patch of the Carter-Penrose diagram of the used eternal black hole geometry is absent in the general gravitational collapse. A further clarification of the mass inflation problem is required beyond the quasi-stationary limit, e.\,g., by using the detailed numerical calculations with the large accretion rate.

\ack

Author acknowledges E.\,O. Babichev, V.\,A. Berezin and Yu.\,N. Eroshenko for the enlightening
discussions. This work was supported in part by the the Research Program OFN-17 of the Division of
Physics, Russian Academy of Sciences.

\end{document}